\documentclass[letter]{aa} % for the letters 
\usepackage{graphicx}
\usepackage{xcolor,colortbl}
\usepackage[]{natbib}
\usepackage{multirow}
\usepackage{amsmath} 
\usepackage{txfonts}  %% txfonts must be after amsmath!!!!!
\usepackage{enumitem}
\usepackage{orcidlink}
\def\degb{^{\circ}}

\begin{document} 

   \title{JWST/MIRI coronagraphic performances as measured on-sky}

   \author{
          A. Boccaletti\orcidlink{0000-0001-9353-2724}\inst{\ref{lesia}},
          C. Cossou\inst{\ref{cea}},
          P. Baudoz\inst{\ref{lesia}},
          P. O. Lagage\inst{\ref{cea}},
          D. Dicken\inst{\ref{UKATC},\ref{Roe}},
          A. Glasse\inst{\ref{UKATC},\ref{Roe}},
          D. C. Hines\orcidlink{0000-0003-4653-6161}\inst{\ref{stsci}},
          J. Aguilar\inst{\ref{stsci}},
          O. Detre\inst{\ref{mpia}},
          B. Nickson\inst{\ref{stsci}},
          A. Noriega-Crespo\inst{\ref{stsci}},
          A. G{\'a}sp{\'a}r\inst{\ref{uarizona}},
          A. Labiano\orcidlink{0000-0002-0690-8824}\inst{\ref{ESAC},\ref{INTA}}, 
          C. Stark\inst{\ref{goddard}},
          D. Rouan\inst{\ref{lesia}},
          J. M. Reess\inst{\ref{lesia}},
          G. S. Wright\inst{\ref{UKATC}},
          G. Rieke\inst{\ref{uarizona}},
          M. Garcia Marin\inst{\ref{esa}},
          }

\institute{
LESIA, Observatoire de Paris, Universit{\'e} PSL, CNRS, Sorbonne Universit{\'e}, Univ. Paris Diderot, Sorbonne Paris Cit{\'e}, 5 place Jules Janssen, 92195 Meudon, France\label{lesia}
\and Université Paris-Saclay, Université Paris Cité, CEA, CNRS, AIM, 91191, Gif-sur-Yvette, France\label{cea}
\and UKATC, The Royal Observatory, Blackford Hill, Edinburgh, EH9 3HJ, Scotland\label{UKATC}
\and  Institute for Astronomy, University of Edinburgh, Royal Observatory, Blackford Hill, Edinburgh EH9 3HJ\label{Roe}
\and Space Telescope Science Institute, 3700 San Martin Dr, Baltimore, Maryland 21218\label{stsci}
\and Max-Planck-Institut für Astronomie (MPIA), Königstuhl 17, 69117, Heidelberg, Germany \label{mpia}
\and Steward Observatory and the Department of Astronomy, The University of Arizona, 933 N Cherry Ave, Tucson, AZ, 85721, USA \label{uarizona}
\and Telespazio UK for the European Space Agency, ESAC, Camino Bajo del Castillo s/n, 28692 Villanueva de la Ca\~nada, Spain\label{ESAC}
\and Centro de Astrobiolog\'ia (CSIC-INTA), Carretera de Ajalvir, 28850 Torrej\'on de Ardoz, Madrid, Spain \label{INTA}
\and NASA Goddard Space Flight Center, Exoplanets \& Stellar Astrophysics Laboratory, Code 667, Greenbelt, MD 20771, USA \label{goddard}
\and European Space Agency, 3700 San Martin Drive, Baltimore, MD21218\label{esa}
}

%   \date{Received September 15, 1996; accepted March 16, 1997}

  \abstract
   {Characterization of directly imaged exoplanets is one of the most eagerly anticipated science functions of the James Webb Space Telescope. MIRI, the mid-IR instrument has the capability to provide unique spatially resolved photometric data points in a spectral range never achieved so far for such objects.}
   {We aim to present the very first on-sky contrast measurements of the MIRI's coronagraphs. In addition to a classical Lyot coronagraph at the longest wavelength, this observing mode implements the concept of the four quadrant phase mask for the very first time in a space telescope.}
   {We observed single stars together with a series of reference stars to measure raw contrasts as they are delivered on the detector, as well as reference subtracted contrasts.}
   {MIRI's coronagraphs achieve raw contrasts greater than $10^3$ at the smallest angular separations (within $1''$) and about $10^5$ further out (beyond $5\sim6''$). Subtracting the residual diffracted light left unattenuated by the coronagraph has the potential to bring the final contrast down to the background and detector limited noise floor at most angular separations (a few times $10^4$ at less than  $1''$). }
   {MIRI coronagraphs behave as expected from simulations. In particular the raw contrasts for all four coronagraphs are fully consistent with the diffractive model. Contrasts obtained with subtracting reference stars also meet expectations and are fully demonstrated for two four quadrant phase masks (F1065C and F1140C). The worst contrast, measured at F1550C, is very likely due to a variation of the phase aberrations at the primary mirror during the observations, and not an issue of the coronagraph itself. We did not perform reference star subtraction with the Lyot mask at F2300C, but we anticipate that it would bring the contrast down to the noise floor.  }


\authorrunning{A. Boccaletti et al.}
\titlerunning{JWST/MIRI coronagraphic performance on-sky}

   \maketitle
%
%-------------------------------------------------------------------

%----------------------------
\begin{table*}[th!]
\begin{center}
\begin{tabular}{lllllllll}
\hline
date      & filter & object   &  type     & obs id  & N$_{group}$ & N$_{int}$ & dither & T$_\mathrm{exp}$                          \\
 UT       &        &          &           &         &             &           &        &  per dither (s) \\  
\hline
\hline
06/09/2022  & F1140C & BD\,+30 2990 & REF 1 ON       & 1037 / obs 4  & 100 & 21 & 9 & 508.122 \\
06/09/2022  & F1140C & HD\,158165   & TARG ON       & 1037 / obs 5  & 100 & 21 & 9 & 508.122 \\
06/09/2022  & F1140C & HD\,158165   & TARG OFF      & 1037 / obs 6  & 10  & 25 & 4 & 65.672 \\
06/09/2022  & F1140C & HD\,158896   & REF 10 ON      & 1037 / obs 7  & 100 & 21 & 9 & 508.122 \\
06/13/2022  & F1140C & $-$          & BGD           & 1045 / obs 65 & 100 & 21 & 4 & 508.122 \\
06/20/2022  & F1140C & BD\,+30 2990 & REF 1 ON 2nd   & 1037 / obs 30 & 100 & 21 & 9 & 508.122 \\
\hline
\hline
06/18/2022  & F1550C &  $-$         & BGD           & 1037 / obs 8  & 100 & 95 & 4 & 2299.49 \\
06/18/2022  & F1550C &  HD\,163113  & TARG OFF      & 1037 / obs 9  & 10  & 25 & 4 & 65.672 \\
06/18/2022  & F1550C &  HD\,163113  & TARG ON       & 1037 / obs 10 & 100 & 95 & 9 & 2299.49 \\
06/18/2022  & F1550C &  HD\,162989  & REF 1 ON      & 1037 / obs 11 & 100 & 95 & 9 & 2299.49 \\
\hline
\hline
06/20/2022  & F1065C & $-$          & BGD           & 1037 / obs 26 & 100 & 21 & 4 & 508.122 \\
06/20/2022  & F1065C & HD\,158165   & TARG OFF      & 1037 / obs 27 & 10  & 25 & 4 & 65.672 \\
06/20/2022  & F1065C & HD\,158165   & TARG ON       & 1037 / obs 28 & 100 & 21 & 9 & 508.122 \\
06/20/2022  & F1065C & BD\,+30 2990 & REF 1 ON       & 1037 / obs 29 & 100 & 21 & 9 & 508.122 \\
\hline
\hline
06/23/2022  & F2300C & $-$         & BGD            & 1037 / obs 32  & 100 & 90 & 4 & 2944.836 \\
06/23/2022  & F2300C & HD\,163113  & TARG OFF       & 1037 / obs 33  & 10  & 25 & 4 & 88.776 \\
06/23/2022  & F2300C & HD\,163113  & TARG ON        & 1037 / obs 34  & 100 & 90 & 1 & 2944.836 \\
\hline
\hline
\end{tabular}
\end{center}
\caption{Main parameters of the observations during MIRI coronagraph's commissioning: date, filter, name of the object, type of object (target or reference, degrees separation on-sky from the target, on or off the center of the coronagraph, or background image), id of the program, number of group, number of integration, number of dither position (9 is for the SGD, 4 or 1 is a classical dither), total exposure time per dither.} 
\label{tab:log}
\end{table*}
%----------------------------

\section{Introduction}

Exoplanet characterization is entering a new era with the James Webb Space Telescope ({\it JWST}), in particular with the coronagraphs of the MIRI instrument \citep{Rieke2015, Wright2015} designed to obtain high contrast imaging of exoplanetary systems at mid-IR wavelengths. To date, several exoplanets have been directly imaged at near-IR, mostly from the ground with adaptive optics facilities, but no observations have been obtained beyond $\sim5\,\muup$m. The few observations performed with ground-based instruments at the M band are potentially affected by large photometric uncertainties due to the sky brightness and variability.

The MIRI coronagraphic mode is a suite of four focal plane masks (permanently mounted in the imager field of view), each paired with a dedicated filter and an optimized Lyot stop. They were designed to offer large contrasts, and, importantly, small inner working angles (IWA)\footnote{The IWA, although sometimes ill-defined, is  the angular separation at which an off-axis point source will have its transmission reduced to 50\%.} at mid-IR. 
Despite longer operating wavelengths than NIRCAM, MIRI's coronagraphs are delivering similar IWAs.
Three of these {coronagraphs} use four-quadrant phase masks \citep[4QPM,][]{Rouan2000}, manufactured with reactive ion etching in a  Germanium substrate. Details about the 4QPM manufacturing for MIRI can be found in \citet{Baudoz2006}. Since the 4QPMs are chromatic, they are used in conjunction with the narrow band filters F1065C ($\lambda_0=10.575 \muup$m, $\Delta\lambda=0.75\muup$m), F1140C ($\lambda_0=11.30 \muup$m, $\Delta\lambda=0.8\muup$m), and F1550C ($\lambda_0=15.50 \muup$m, $\Delta\lambda=0.9\muup$m), mounted in the filter wheel together with a Lyot stop transmitting 62\% of the telescope aperture. Because the stop is not just a downsized version of the JWST pupil but has been optimized to attenuate the starlight diffraction, the PSF of an off-axis object is slightly broader than with the full pupil. The field of view (FOV) of the 4QPM subarrays is $24''\times24''$. 
Another coronagraph using a classical Lyot mask with a diameter of $3\lambda/D$, representing $2.1''$, is dedicated to longer wavelength observations with the F2300C filter ($\lambda_0=22.75 \muup$m, $\Delta\lambda=5.5\muup$m). This coronagraph, with a FOV of $30''\times30''$, uses a Lyot stop with a different shape and a transmission of $72\%$. Pre-flight details about the MIRI coronagraphs can be found in \citet{Boccaletti2015}.

The central wavelengths of the filters were chosen to characterize the atmospheres of directly imaged, young giant exoplanets, essentially to complement near-IR photometric and spectroscopic measurements. The F1065C and F1140C filters are meant to provide photometric measurements of exoplanets in and out of the ammonia absorption band. 
The F1550C filter, in combination with F1140C, is important to constrain models of the thermal balance of the planet and of the behaviour of any atmosphere.
%The F1550C, in combination with F1140C, is considered a good proxy for estimating a planet's temperature.
The detection performance was first estimated in \citet{Boccaletti2005}, \citet{Boccaletti2015}. Some recent simulations of planet detection with MIRI's coronagraphs can also be found in \citet{Danielski2018} and \citet{Hinkley2022}. The Lyot mask, by design, is not intended to directly image planets (except a few at wide separations like HD\,106906b), but rather is optimized to detect cold material around bright stars, such as debris disks, which are reminiscent of Kuiper belts in young systems \citep[see for instance][]{Lebreton2016}. The larger FOV in this channel was adopted to include so far as possible the outer debris belts of even the most extended systems.

The MIRI coronagraphs were previously tested on the ground but the performances were strongly disturbed by the strong thermal background and the test bench features, which limit the achievable contrast to a few hundreds \citep{Cavarroc2008spie}. 
The purpose of this paper is to present the actual performance of the MIRI coronagraphs as measured during JWST commissioning, and to compare these results to simulations, which were recently revised to include the most relevant knowledge of the telescope and the instrument.
 
The paper is organized as follows: Section \ref{sec:obs} describes the observations, and Section \ref{sec:simul} the simulations. The measured contrasts on-sky are presented in Section \ref{sec:contrasts}.

\begin{figure*}
     \centering
        \includegraphics[trim = 0cm 0cm 0cm 0cm, width=\textwidth]{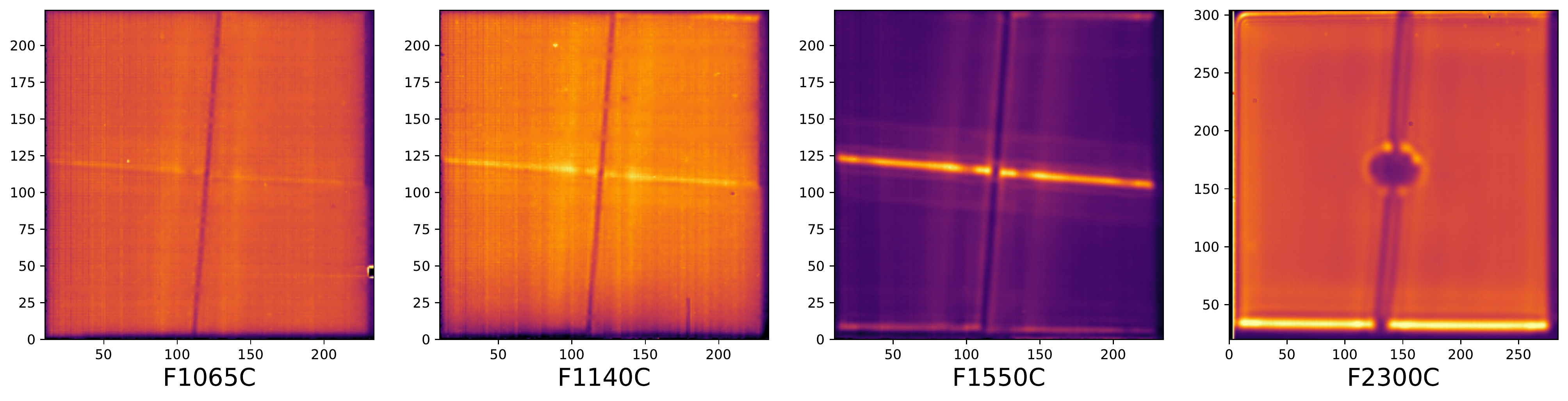}
      \caption{Images of the background obtained in the coronagraphic subarrays showing the "glow sticks" visible at the horizontal transitions of the 4QPMs, and around the disk of the Lyot mask, but also at the edges of the coronagraphs' support structure. The intensity scale is adapted for each image for visualization purpose.}
\label{fig:glowstick}
\end{figure*}

%----------------------------
\begin{table*}[th!]
\begin{center}
\begin{tabular}{lllllllll}
\hline
    &   OPD tel.
    & OPD MIRI  & OPD Frill  & OPD IEC   & OPD TD   & TA  &   jitter  & pupil shear \\
    &   [nm]      & [nm]  & [nm]   & [nm]  & [nm] & [mas/axis] & [mas/axis] & [\%] \\
\hline
\hline
best        & 73 & 32  & 0.017 & 1.6 & 0.016 & 6.25  & 2.5 & 0 \\
nominal     & 73 & 32  & 0.07  & 2.8 & 0.07  & 8.76  & 3.8 & 2 \\
requirement & 73 & 32  & 1.75  & 4.6 & 1.67  & 12.51 & 5.8 & 4 \\
\hline
\hline
\end{tabular}
\end{center}
\caption{Three input cases used to simulate the performance of the MIRI coronagraphs. Wavefront errors (OPD) are provided in nanometers RMS, TA and jitter are in milliarcsec per axis, and the shear is in \% of the telescope pupil.} 
\label{tab:param_simul}
\end{table*}
%----------------------------

%--------------------------------------------------------------------
\section{Observations}
\label{sec:obs}
Observations were carried out in June 2022. Table \ref{tab:log} provides the list of data that are used for estimating the contrasts delivered by the MIRI coronagraphs. 
We selected the targets to avoid saturation when observed off-axis (a few arcsec away to avoid the attenuation of the coronagraphs) to measure the PSF for a photometric reference, as well as to provide a sufficient flux level compatible with the sensitivity at each filter when the star is on-axis, hence masked and attenuated by the coronagraph (attenuation on-axis can be as large as a factor of
$\sim$400 as measured in F1550C). We end up with two series of objects, one set optimized for F1065C and F1140C, and another one for F1550C and F2300C. The targets are HD\,158165 (K=4.07) of which the flux densities are 0.550\,Jy and 0.450\,Jy, for respectively F1065C and F1140C, and then HD\,163113 (K=2.75), for F1550C and F2300C, with flux densities 1.445\,Jy and 0.532\,Jy, respectively.

The observations recorded during the commissioning of MIRI coronagraphs are intended primarily to measure the raw contrasts, which correspond to the starlight attenuation at the detector as a function of angular separation. Raw coronagraphic images in this particular case are entirely dominated by the diffraction of the telescope pupil leaking through the Lyot stop. The un-attenuated starlight can be further calibrated and subtracted with the use of reference stars.
At this second step, the dominant terms are expected to come from the telescope wavefront errors and their variations in time, as well as the telescope pointing repeatability onto the coronagraph. 
While a long-term strategy will certainly involve a library of coronagraphic images and dedicated algorithms for post-processing \citep{Choquet2014}, the commissioning procedure makes use of one or two reference stars observed back-to-back with the targets, or a few days apart.  

Reference stars were chosen to have similar flux densities in the coronagraphic filters, and are located at several distances on the sky to test the effect of telescope slews. The 4QPM/F1140C coronagraph is the one identified to perform most of these tests with reference stars. In the F1140C filter, we observed two reference stars, BD\,+30 2990 and HD\,158896, located respectively at 0.7$^\circ$ and 20.7$^\circ$ from the target. The former was also observed in the F1065C filter. In the F1550C filter, the reference star HD\,162989 is at 1.16$^\circ$ from the target. No reference star was observed for the Lyot coronagraph.

%Because of unexpected features, referred as the "glow sticks", materializing at the edges of the focal plane components (in particular the horizontal transitions of the 4QPMs, and the edges of the Lyot mask), the initial procedure to locate the position of the nulls of the coronagraphs had to be adapted in flight.

Figure \ref{fig:glowstick} shows images of each of the coronagraphic fields when observing the background sky.  All four show an unexpected stray light feature, known familiarly as 'glow sticks', which appear as increased signal along the structural edges in the MIRI imager entrance focal plane. 
%that are approximately parallel to the detector row direction.  
They are most apparent across the centre of the F1550C image where light is being scattered into the 'science' optical path by the raised edge of the phase boundary.  In the F2300C image, the bright glow stick marks scattering at the lower edge of the aluminium Lyot aperture. The stray light is visible, but fainter for the shorter wavelength 4QPM coronagraphs in Figure \ref{fig:glowstick}. Stray light is also seen along the upper edge of the Lyot spot.

Two key observations were important in determining the root cause of the glow sticks.  First, their shape and brightness were independent of the observatory pointing direction to within a few percent, ruling out an astronomical origin.  Second, photometric analysis of the glow sticks determined that the illuminating source was well fitted by a grey body spectrum with an effective temperature of $120 \pm 20 K$, characteristic of the region where the sunshield approaches the deployable tower assembly \citep{Lightsey2012}.  Non-sequential optical path analysis (S. Rohrbach, private com.) has identified a possible path from this warm region to the MIRI entrance focal plane via a reflection from the secondary mirror (SM) followed by scattering from the hinged SM support strut.  Although the cosmetic appearance of the modelled glow sticks is not identical to those observed, with the along column edges also predicted to be bright, we regard it as sufficiently close to be consistent with the observations within the fidelity of the observatory plus MIRI solid model and the as-built reality.

To mitigate the 'glow stick' effect (which could be brighter than the observed source itself) it is necessary to subtract a background image obtained in the same filter and for identical exposure time, until the variability of this pattern is understood and an alternative  approach is proposed. If this procedure is followed, the effect of the glow sticks on the final data is virtually completely removed except for the expected modest increase in photon noise at their positions.

%An unexpected feature was observed in the coronagraphic images, referred to as the "glow sticks", materializing at the edges of the focal plane components (in particular the horizontal transitions of the 4QPMs, and the edges of the Lyot mask). The glow sticks are particularly obvious in the background images as shown in Fig. \ref{fig:glowstick}. While undetected at the shortest wavelengths, they start to kick off at about 10\,$\muup$m, right at the operating wavelength of the coronagraphs, and are getting stronger at longer wavelengths. The origin of the glow sticks is not fully established but seems to come from a thermal emission between the primary mirror and the sun shield, which reflects on the secondary mirror and its struts, and finally makes it way into the science path. Therefore, we recommend that each coronagraphic observation should include an observation of the background to be subtracted, with a comparable signal as the on-source observation, until the variability of these patterns is understood and possibly calibrated.

To precisely center the star on the coronagraph axis, we need both sub-pixel estimation of the coronagraph position and a precise target acquisition procedure \citep{Cavarroc2008}. The glow sticks prevented us from using the dark transitions of the 4QPMs to estimate the coronagraph center, as planned. Instead, we developed a method based on comparison with a diffraction model after subtracting them out, to be presented in a separate paper (Baudoz et al., in prep.). These measurements were used in the target acquisition procedure to reach a pointing accuracy of about 5 to 10\,mas in full agreement with the requirement \citep{Rigby2022}.
%\dean{[\bf{Jame Rigby's commissioning document released publicly says 10 mas, which I think she got from Alistair's readiness document.}]}

%Instead of measuring the dark transition of the 4QPMs, as proposed in REF, we developed a method based on comparison with a diffraction model, to be presented in a separate paper (Baudoz et al., in prep.), and with which we demonstrated an accuracy of better than 5\,mas. These measurements were used in the target acquisition procedure.

\begin{figure*}
     \centering
        \includegraphics[trim = 0cm 0cm 0cm 0cm, width=\textwidth]{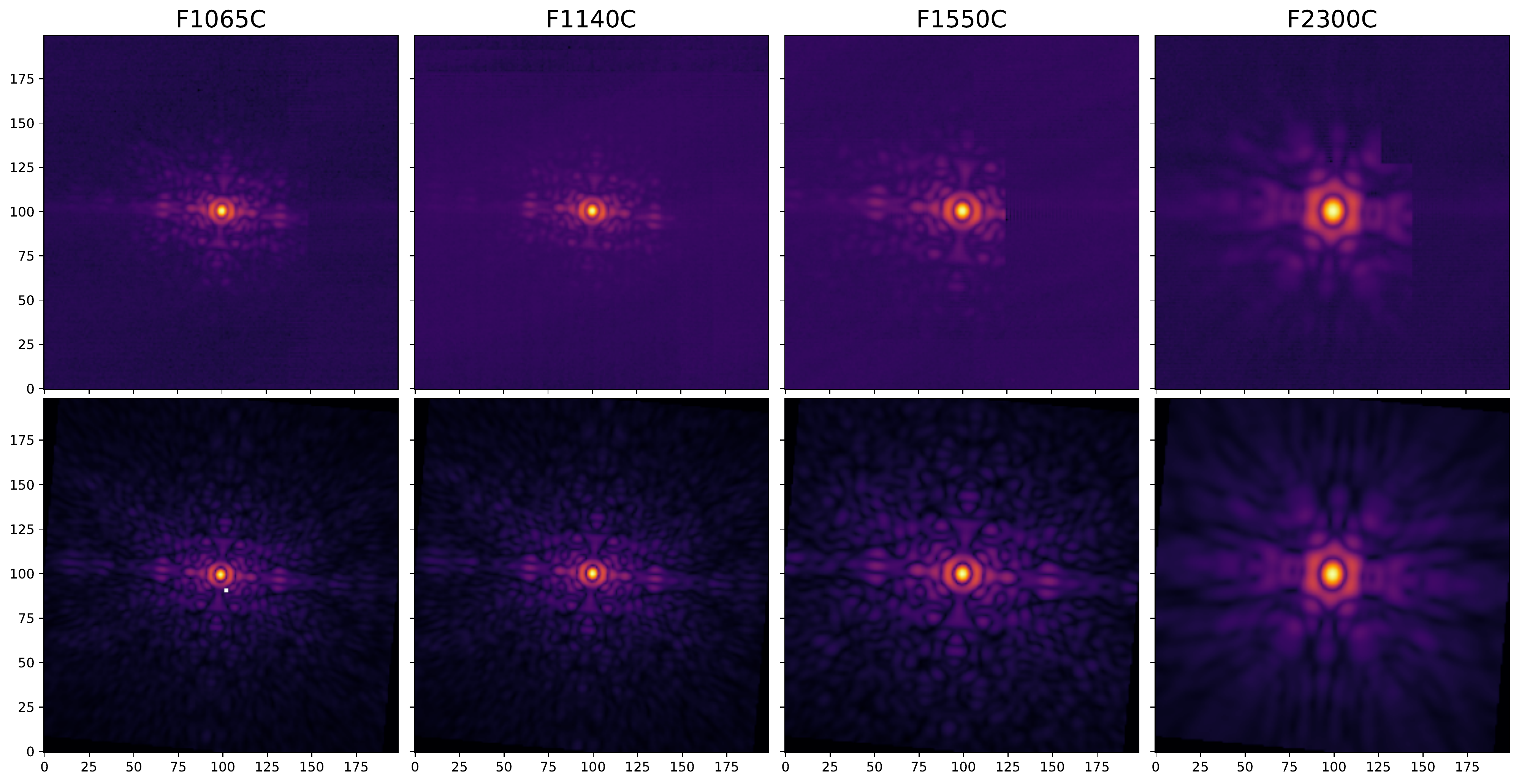}
      \caption{Observed (top) versus simulated (bottom) PSFs (off-axis) in the four coronagraphic filters. The pixel scale is 110\,mas. Some PSFs are cropped by the edges of the coronagraph's mechanical support. Simulated data are using the nominal scenario (see sec. \ref{tab:param_simul})}
\label{fig:psf}
\end{figure*}

\begin{figure*}
     \centering
        \includegraphics[trim = 0cm 0cm 0cm 0cm, width=\textwidth]{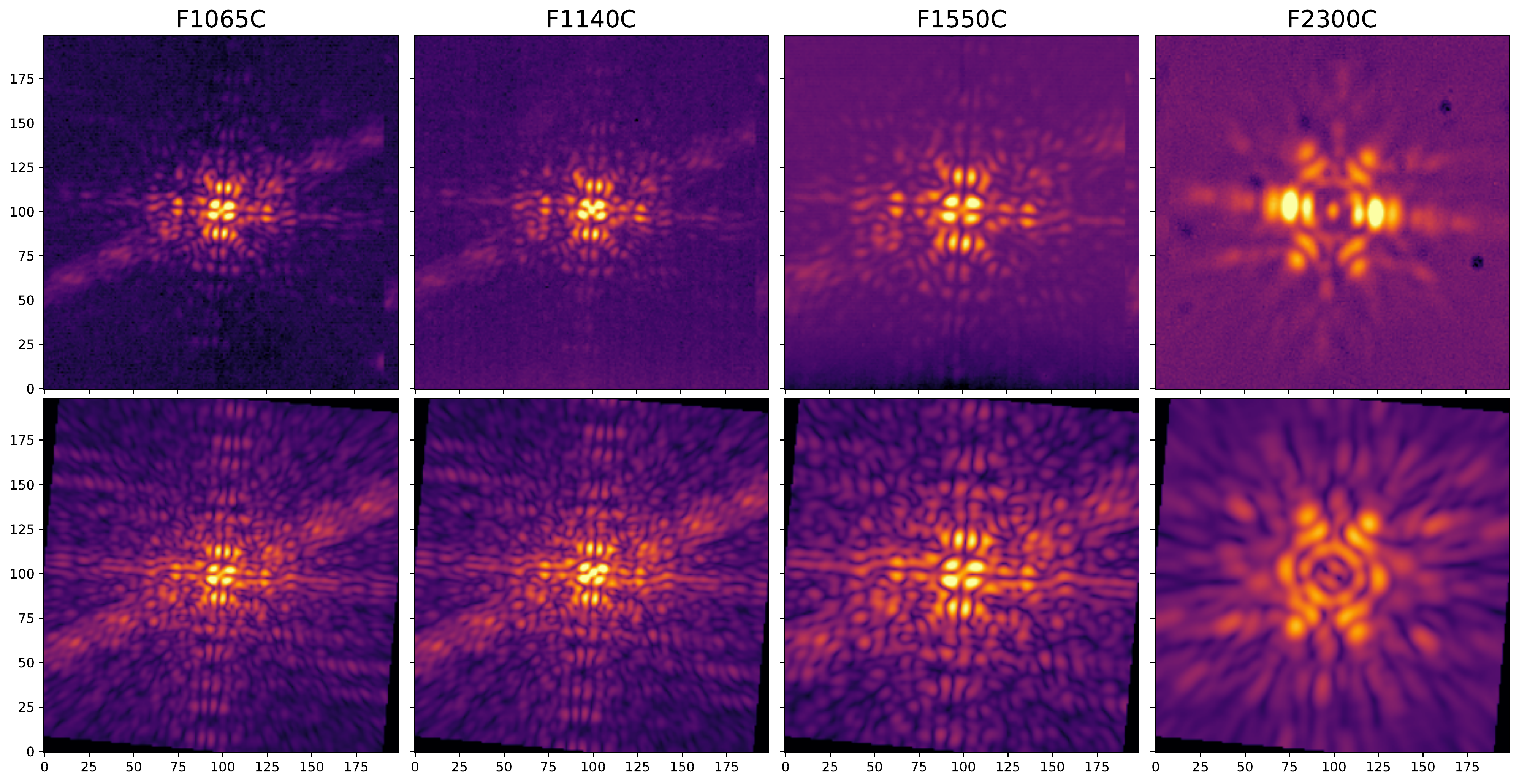}
      \caption{Observed (top) versus simulated (bottom) coronagraphic images (on-axis) in the four coronagraphic filters. The pixel scale is 110\,mas. Dark spots in the F2300C image corresponds to background stars recorded during the background observation. Simulated data are using the nominal scenario (see sec. \ref{tab:param_simul})}
\label{fig:raw}
\end{figure*}

\begin{figure*}
     \centering
        \includegraphics[trim = 0cm 0cm 0cm 0cm, width=\textwidth]{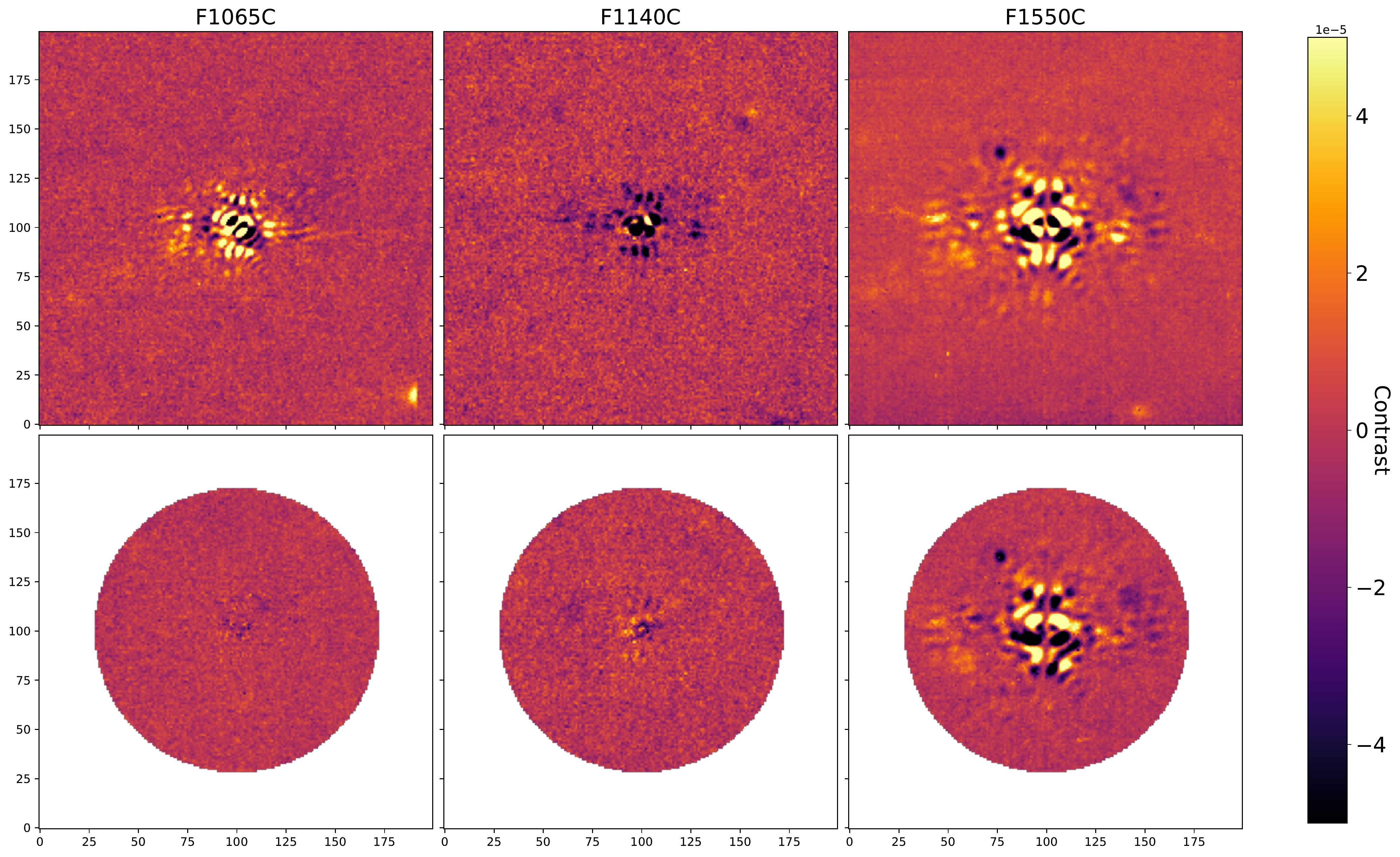}
      \caption{One to one (top) versus PCA (bottom) subtraction of coronagraphic images (TARG and REF1) in the three 4QPMs filters. Several background sources (either positive or negative) are observed around the target and reference stars. The pixel scale is 110\,mas. }
\label{fig:sub-pca}
\end{figure*}

\section{Simulations}
\label{sec:simul}
The simulations presented in \citet{Boccaletti2015} were recently reassessed by the {\it JWST} Coronagraph Sensitivity Working Group to incorporate the up-to-date telescope and instrument parameters. We assumed a temporal sampling of $\pi/5\approx 0.63$ minutes per frame, for a total sequence of about 56 minutes on the target (90 frames) followed by the same amount of telescope time on a reference star, which is dithered on 9 positions ($9\times10$ frames, i.e., each dither position has a total of 1/9 the exposure time as the target source.). 
This so-called Small Grid Dither (SGD) allows a diversity in the observations of the reference star to further reduce the starlight making use, for instance, of Principle Component Analysis \citep{Lajoie2016, Soummer2012}. The SGD is a square grid with 10\,mas steps. Although the error on the positioning of the star onto this grid is estimated to 1 or 2\,mas, this has no effect in the estimation of the starlight which only relies on the variations of the intensity of speckles around the mask center, but not on absolute knowledge of the pointing.

Static aberrations include the telescope wavefront aberration as measured in the early phase of commissioning (73\,nm RMS), and the MIRI instrument aberrations measured on the ground (32\,nm RMS). The former wavefront map is made of mid spatial frequencies and is expected to evolve along the life of the mission, while the latter contains mostly low spatial frequencies. Additional dynamical components in the wavefront on a $\sim1-2$ hours timescale are also taken into account with various spatial and temporal frequencies like the Thermal Distortion (TD) of the telescope backplane, the fast oscillation in the heaters in the ISIM Electronics Compartment (IEC), and the Frill around the primary mirror designed to stop the straylight, all being relatively small in terms of wavefront errors for an instrument like MIRI (Tab. \ref{tab:param_simul}). 

In addition, the simulation accounts for misalignments at the focal plane and pupil plane in the coronagraph. First of all, the offset between the star's position, and the center of the mask corresponds to the Target Acquisition (TA) error. For convenience, in the simulations, the target star is perfectly centered while the reference star is offset by this TA error. Then, we included line of sight jitter that is the motion of the star's position during the observation. Moreover, the Lyot stop located at the MIRI filter wheel, can be slightly misaligned with the telescope pupil. This error is expressed in \% of the telescope pupil diameter, assuming a shear along the diagonal. Three distinct scenarios were considered: 'best', 'nominal' and 'requirement'. The values for all parameters of the simulations are provided in Tab. \ref{tab:param_simul}. 
Finally, we included a spectral shift of 3\% of the F1140C filter with respect to the operating wavelength of the corresponding 4QPM, causing a chromatic leakage in the images visible as a central peak in the coronagraphic image. 

The simulations are time averaged, so we are left with one single image for the target and nine for the reference. We applied PCA using 9 components to built a reference frame then subtracted out from the target image. We provide a comparison of PSF images (off-axis) and coronagraphic images (on-axis) actually observed during commissioning, to simulated images in Fig. \ref{fig:psf} and Fig. \ref{fig:raw}.
They are virtually identical and the surrounding field on the observed image is very clean, with no evidence of residual latent images.

%-------------------------------------- Two column figure (place early!)
 
\begin{figure*}
     \centering
        \includegraphics[trim = 5cm 2cm 4cm 2cm,width=0.45\textwidth]{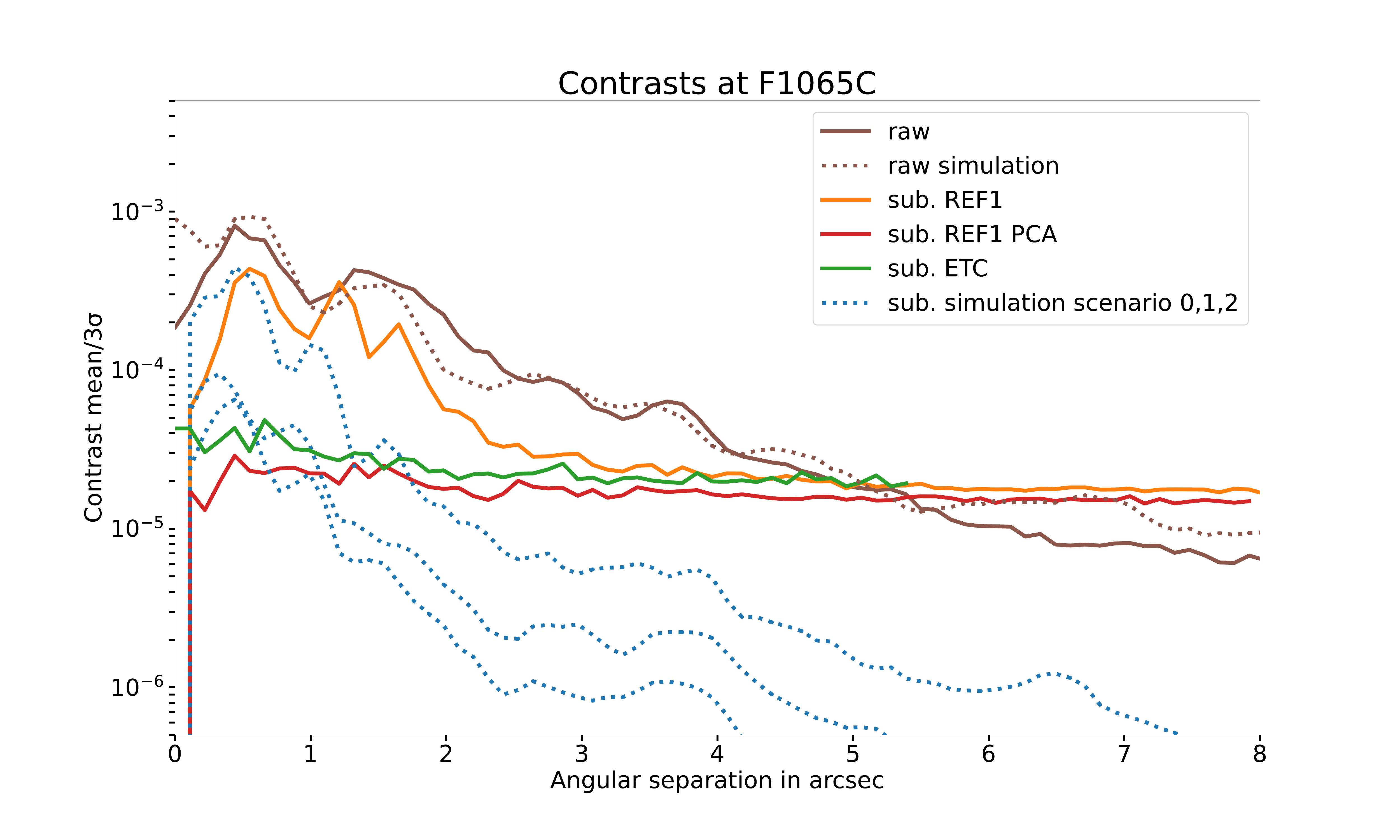}
        \includegraphics[trim = 4cm 2cm 5cm 2cm,width=0.45\textwidth]{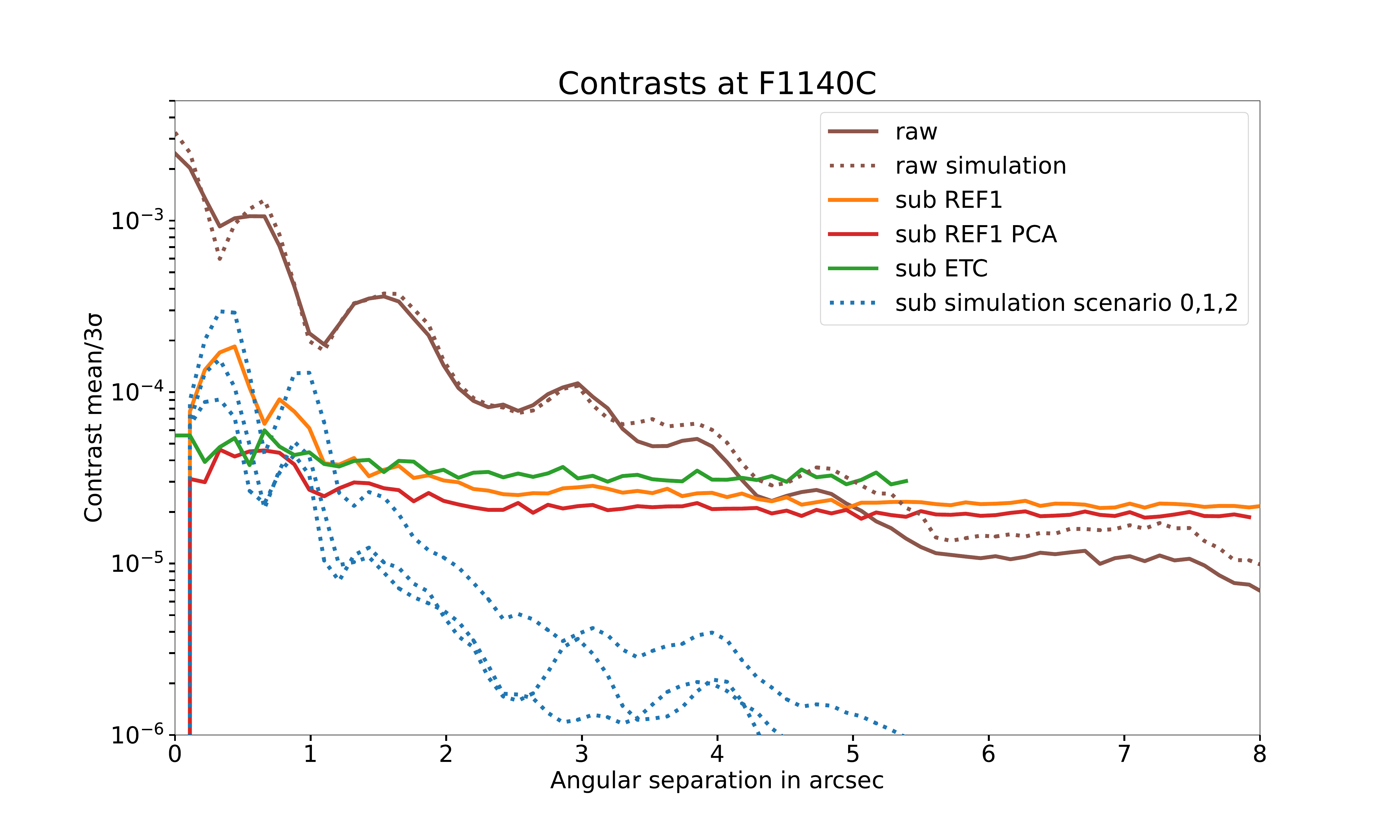}
        \includegraphics[trim = 5cm 2cm 4cm 2cm,width=0.45\textwidth]{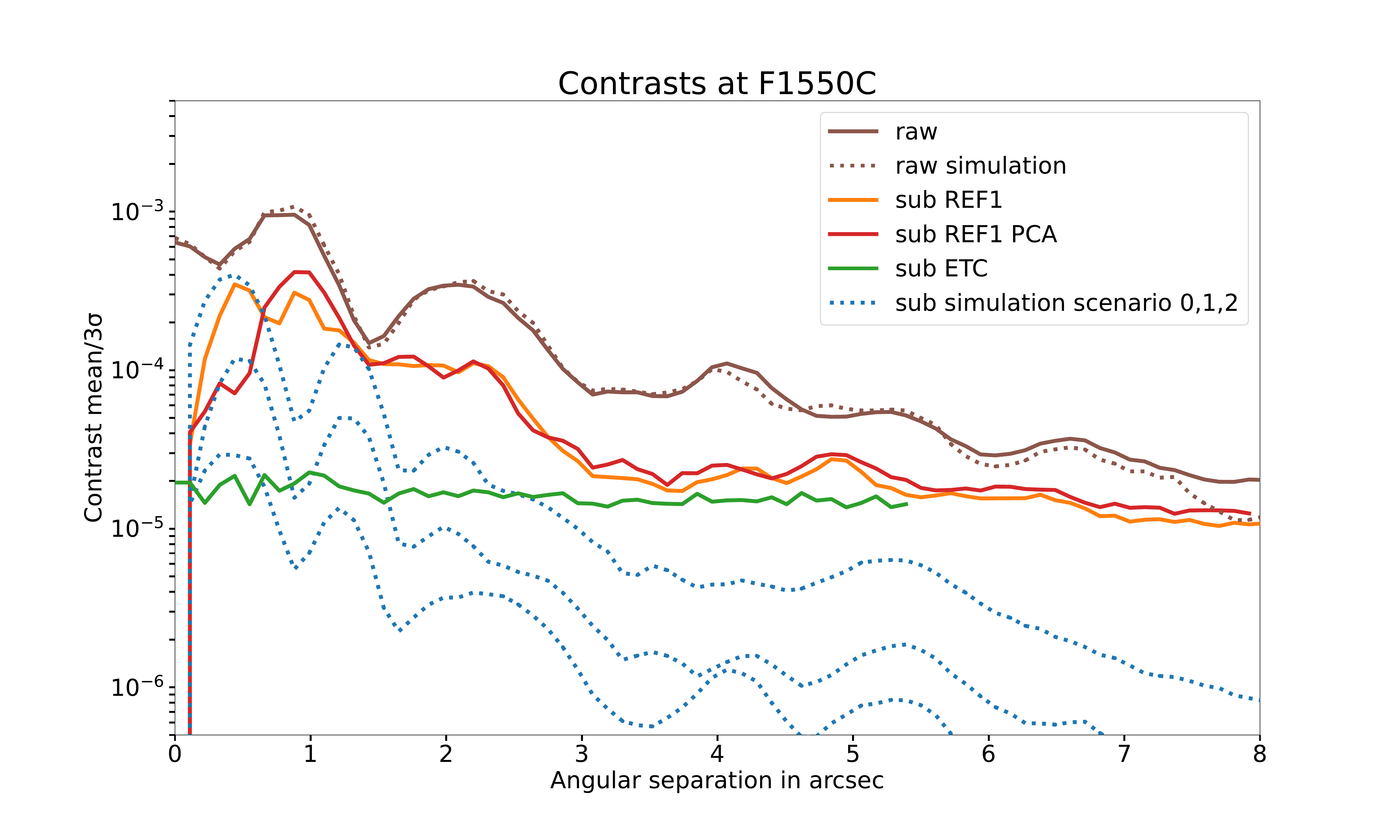}
        \includegraphics[trim = 4cm 2cm 5cm 2cm,width=0.45\textwidth]{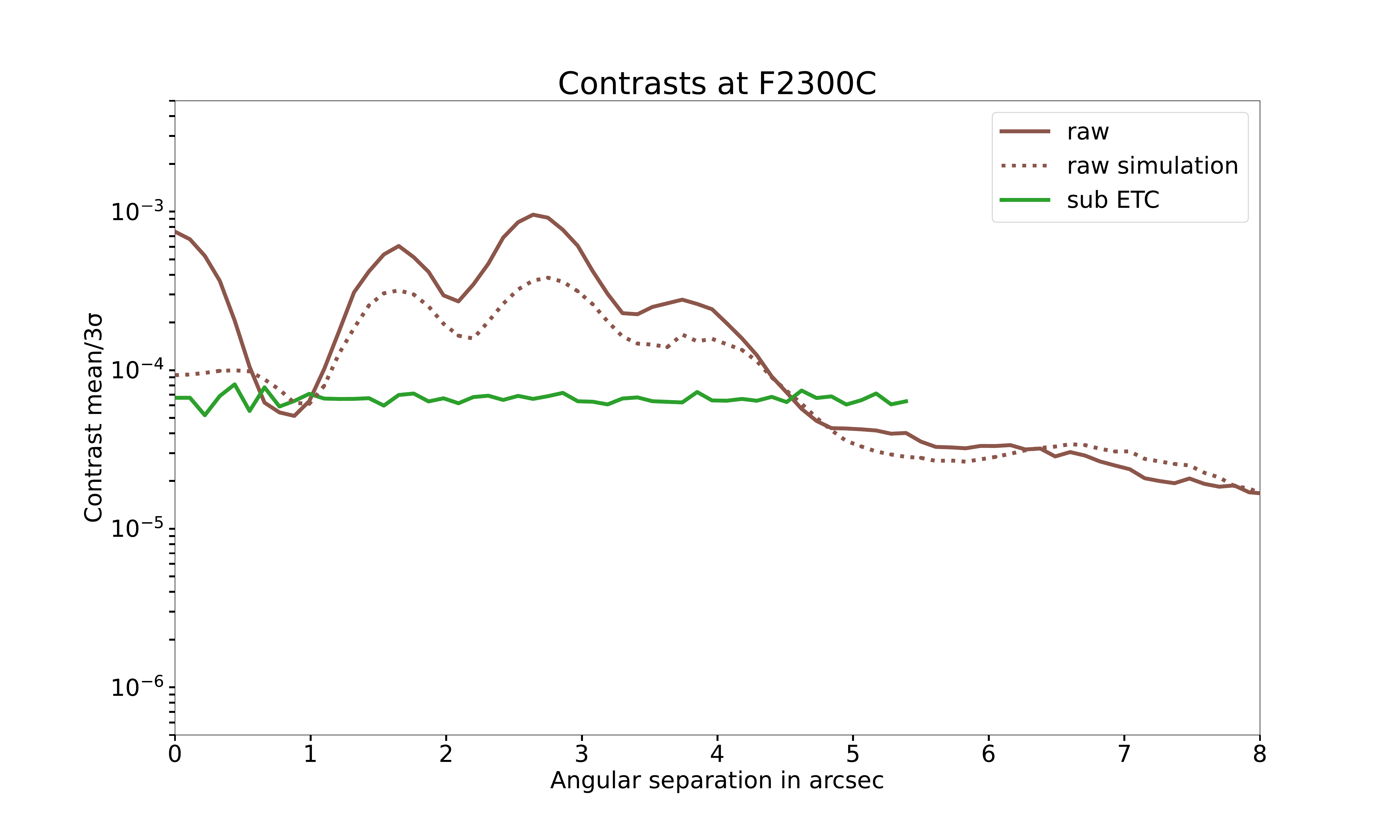}
     \caption{
    Measured raw and 3$\sigma$ contrasts for all MIRI coronagraphic filters as compared to the simulations and to the ETC prediction. Subtracted contrasts are shown for the one to one and PCA algorithms. Reference stars were not observed at F2300C. 
     }
     \label{fig:contrasts}
\end{figure*}

\begin{figure}
     \centering
     \includegraphics[trim = 4cm 2cm 5cm 2cm,width=0.45\textwidth]{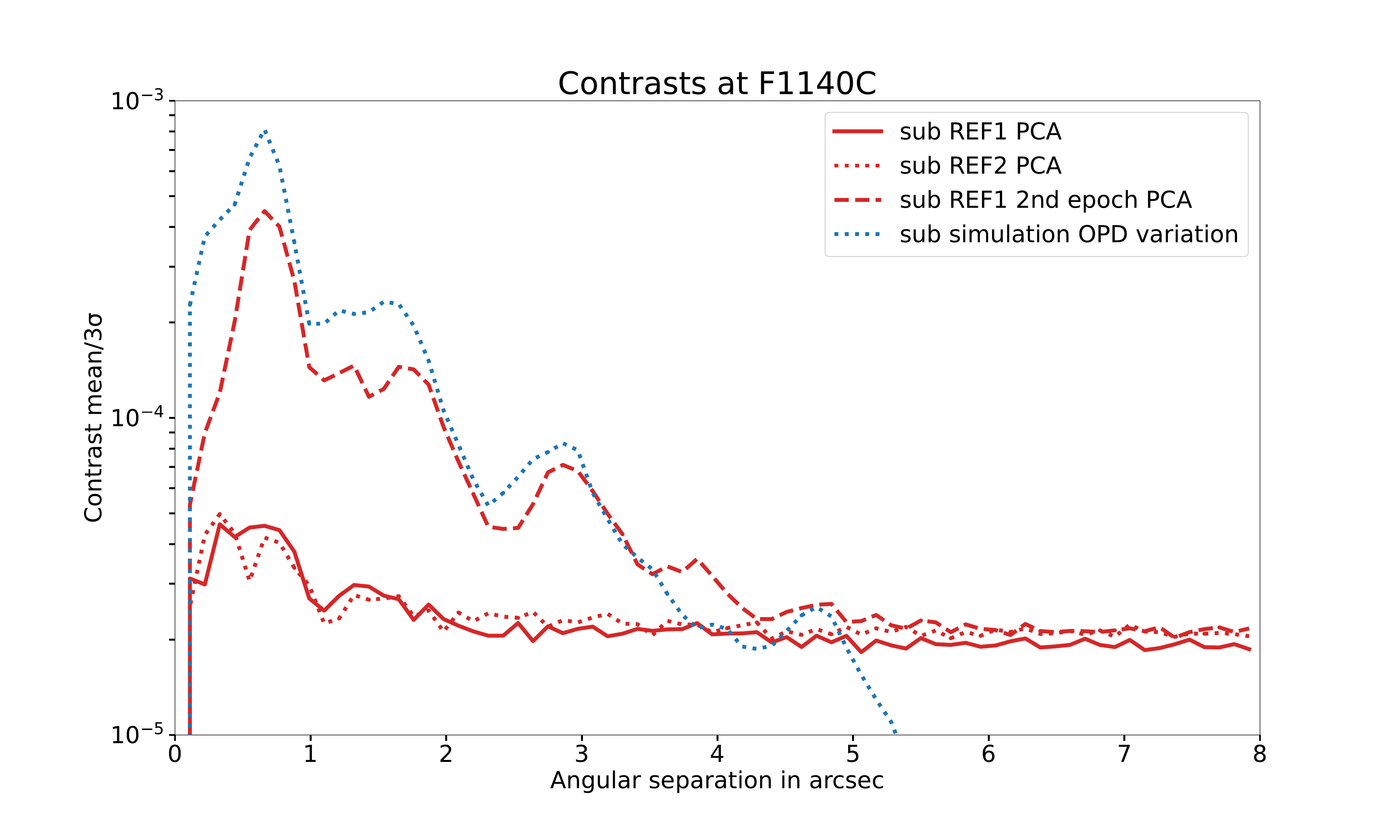}
     \caption{Measured 3$\sigma$ contrasts with PCA in the F1140C filter for the single epoch and the two epochs cases, and compared to the simulation taking into account the wavefront error maps (OPD) before and after rephasing of the primary mirror segments.}
     \label{fig:contrasts1140}
\end{figure}

\section{Measured contrasts}
\label{sec:contrasts}
The data are processed with the dedicated {\it JWST} pipeline up to level 2 (but without Flat Field correction), then we used our own specific procedures for subtracting reference star images.
Contrasts are measured azimuthally as a function of the angular separations first in the raw coronagraphic image (azimuthal mean), and then with the subtraction of a reference image (azimuthal standard deviation). The reference-subtracted image is created either by subtracting a single reference exposure from the target exposure (one-to-one subtraction), or by subtracting an image constructed from a combination of multiple reference exposures (PCA subtraction).
In commissioning we used the SGD mode for both the target and the reference, so we have 9 images for each, which means 81 possible  one-to-one subtractions and 9 PCA subtractions. We only display a resulting contrast curve which takes the best value of the contrasts for each separation, independently (consistent with the best subtraction producing the largest contrast at all separations). 
%OK mais c'est un peu de la triche par rapport aux observations qui suppose seulement 1 observation de l'étoile et 9 observations de la ref, non ? 
Examples of the subtracted images for the three 4QPMs are given in Fig. \ref{fig:sub-pca}.
The measurements are compared to the noiseless simulated contrasts for the three scenarios mentioned in section \ref{sec:simul}, and to the ETC estimations using the actual flux of the observed stars and exposure times, together with a "medium background" configuration. The ETC assumes that the reference star subtraction is limited by photon and detector noises, so does not capture a possible variability in the optical wavefront aberrations, or pointing.  

Figure \ref{fig:contrasts} displays the contrast curves for all four coronagraphic filters. The raw contrasts (brown lines) are in good agreement with the  simulations of the 4QPMs coronagraphs. The F1065C and F1550C show the characteristic dip in the center, while the central region with F1140C is clearly affected by the chromatic leakage. Raw contrasts are usually better than $\sim10^{-3}$, and reach the background limit at $\sim10^{-5}$ (beyond 6$''$ or so). The one-to-one subtraction (orange lines) only provides a small gain with respect to the residual diffraction left behind the coronagraph, while PCA (red lines) offers a more substantial attenuation, reaching the background floor at all angular separations in F1065C and F1140C, which is in line with ETC expectations (green lines). 
The two references (REF1 and REF2 in Fig. \ref{fig:contrasts1140}) observed in F1140C back to back with the target produce very comparable contrast limits (dotted red line) indicating a small dependence with telescope slew (at least for a slew amplitude lower than $\sim20^\circ$). In the case we apply PCA using the reference star observed 11 days later (REF1 2nd epoch in Fig. \ref{fig:contrasts1140}) the contrast significantly degrades (dashed red line). We interpret it as a result of telescope re-phasing during this time frame.  At the end of May 2022, a micro meteorite impacted the segment C3, causing a significant departure of the wavefront with respect to the initial state. The first part of the F1140C observations (TARG, REF1 and REF2) were observed in this configuration with approximately 86\,nm RMS of aberrations in total on the primary mirror, while the 2nd epoch of REF1 was obtained after mirror rephasing leading to a reduced amount of 65\,nm RMS aberrations. Such a difference is not visible in the raw coronagraphic images when comparing the 2 epochs. However, the subtracted contrast in PCA if using these 2 epochs is reduced by as large as an order of magnitude at a separation of $0.66''$ compared to the case with a single epoch. The overall contrast is affected up to a distance of about $4''$. We retrieved the wavefront measurements bracketing these data, assigning the first map to the target and the second to the reference, in order to model the loss of contrast. We obtained almost the same contrast curve as the one measured (blue dotted line compared to red dashed line in Fig. \ref{fig:contrasts1140}) giving credit to our hypothesis.

Overall, for F1065C and F1140C, PCA subtraction using reference stars achieves a contrast of 2 to $4.10^{-5}$ inside 1$''$. This is compatible with the best case scenario in the noiseless simulation showcasing the very good performance met by the observatory in terms of pointing, repeatability and stability. In fact, we estimate that the main parameters defining the subtracted contrast are even better than the best case with a line of sight jitter likely in the 1-2\,mas range, and TA at a level of 5\,mas in total. A qualitative comparison with simulated images indicates a pupil shear of about 2-3\% (responsible for the strip at $\sim45\degb$ as seen in the coronagraphic images in Fig. \ref{fig:raw}). This value will be refined for each coronagraph configuration in Baudoz et al. (in prep.).

In the F1550C filter, the raw contrasts are in perfect agreement with the simulation, but both the one-to-one, and the PCA subtraction are much worse than the ETC or simulated predictions. The 3$\sigma$ contrast achieves only $\sim4.10^{-4}$ at $1''$ while we could expect $2.10^{-5}$. Since the starlight rejection by the 4QPMs itself is coherent with the model, the only plausible explanation is again a mismatch in terms of wavefront errors between the target and the reference star. 
In fact, a 'tilt event' (change in segment position) on one of the primary mirror segment has presumably occurred in between the 16th and the 19th of June 2022, and could be the cause of such a reduced contrast. 
The project has committed to making telescope OPDs available on roughly 2-day centers, which could identify if it is plausible that such an event occurred. There is currently no approach to locate the time of a tilt even more accurately. 
%There is currently no immediate solution to test this hypothesis. 

Finally, we obtained raw contrast measurement with the F2300C Lyot mask. We find that the level of contrast matches qualitatively the model prediction but does not agree perfectly. The discrepancy can be as large as a factor of 2 to 3 between 2$''$ and 3$''$  (if we omit distances that are inside the mask hence irrelevant). The observed image itself features a stronger diffraction in the direction perpendicular to the bar holding the Lyot spot, producing an asymmetrical image in contrast to the simulation (Fig. \ref{fig:raw}). The reason of this disagreement is still under investigation, but we can confidently assume that like the other coronagraphs, the Lyot mask performance will be set by the background level when using a reference subtraction. In fact, this is even more the case at F2300C since the background is stronger. Indeed, while the exposure times are similar in F1550C and F2300C, the achievable contrast is $\sim$4 times worse with the Lyot mask (as predicted with ETC).

\section{Conclusion}
Commissioning observations are the first opportunity to actually measure the performance of the MIRI coronagraphs in real conditions, because high contrast imaging was not practical during the ground testing phase.
The pointing accuracy and reproducibility have been proved excellent, meeting the specifications of 5\,mas at best, and definitely less than 10\,mas (1/10th to 1/20th of a pixel).
All four coronagraphs, the 4QPMs and the Lyot, behave satisfactorily on point sources, in the sense that the coronagraphic images and the raw contrasts are almost identical to the models, and meet the contrast specifications. A small difference is found with the Lyot mask, for which an additional diffraction is superimposed on the predicted image, perpendicularly to the bar. 
Calibrating the residual diffraction left unattenuated by the 4QPMs with reference stars brings the contrast to the limit imposed by thermal background and detector noises, at least for the two shortest wavelength filters, F1065C and F1140C. The reference star subtracted contrast with the F1550C is likely limited by variations of the wavefront errors of the primary mirror which may have occurred during the observations (a 'tilt' event). Similarly, subtracting images taken a few days apart shows significant deterioration of the contrast due to primary mirror re-phasing between the two epochs. 
At the moment, it is recommended that each coronagraphic observations with MIRI include: a background image of the same duration as the science exposure, a 9-point small grid dither on the reference star (5-point SGD was not tested during commissioning), and an off-axis image of the star for photometric calibration purpose. Alternatively, one can use the Target Acquisition images for photometry, but these are obtained with different filters (usually a neutral density filter, FND).

The commissioning of the MIRI Coronagraphs reported here has demonstrated excellent performance, that the use of the 4QPM technique provides the expected small inner working angle, and rejection factors and sensitivity in excess of pre-launch expectations. We can therefore anticipate that with the observing recommendations in this paper, the MIRI coronagraphs will have a key role in direct imaging of exoplanets to constrain atmospheric properties for the very first time at mid IR wavelengths.    

\begin{acknowledgement}
The work presented is the effort of the entire MIRI team and the enthusiasm within the MIRI partnership is a significant factor in its success.  We would like to thank Scott Rhorbach for his invaluable assistance and modelling to resolve the cause of the glowsticks issue, and the JWST and MIRI commissioning teams for their support in the execution of MIRI commissioning.

The following National and International Funding Agencies funded and supported the MIRI development: NASA; ESA; Belgian Science Policy Office; Centre Nationale d’Etudes Spatiales (CNES); Danish National Space Centre; Deutsches Zentrum fur Luft-und Raumfahrt (DLR); Enterprise Ireland; Ministerio De Economiá y Competividad; Netherlands Research School for Astronomy (NOVA); Netherlands Organisation for Scientific Research (NWO); Science and Technology Facilities Council; Swiss Space Office; Swedish National Space Board; and UK Space Agency.

 MIRI draws on the scientific and technical expertise of the following organizations: Ames Research Center, USA; Airbus Defence and Space, UK; CEA-Irfu, Saclay, France; Centre Spatial de Li{\`e}ge, Belgium; Consejo Superior de Investigaciones Cientficas, Spain; Carl Zeiss Optronics, Germany; Chalmers University of Technology, Sweden; Danish Space Research Institute, Denmark; Dublin Institute for Advanced Studies, Ireland; European Space Agency, Netherlands; ETCA,  Belgium; ETH Zurich, Switzerland; Goddard Space Flight Center, USA; Institute d’Astrophysique Spatiale, France; Instituto Nacional de T{\'e}cnica Aeroespacial, Spain; Institute for Astronomy, Edinburgh, UK; Jet Propulsion Laboratory, USA; Laboratoire d’Astrophysique de Marseille (LAM), France; Leiden University, Netherlands; Lockheed Advanced Technology Center (USA); NOVA Opt-IR group at Dwingeloo, Netherlands; Northrop Grumman, USA; Max Planck Institut f{\"u}r Astronomie (MPIA), Heidelberg, Germany; Laboratoire d’Etudes Spatiales et d’Instrumentation en Astrophysique (LESIA), France; Paul Scherrer Institut, Switzerland; Raytheon Vision Systems, USA; RUAG Aerospace, Switzerland; Rutherford Appleton Laboratory (RAL Space), UK; Space Telescope Science Institute, USA; Toegepast-Natuurwetenschappelijk Onderzoek (TNOTPD), Netherlands; UK Astronomy Technology Centre, UK; University College London, UK; University of Amsterdam, Netherlands; University of Arizona, USA; University of Bern, Switzerland; University of Cardiff, UK; University of Cologne, Germany; University of Ghent; University of Groningen, Netherlands; University of Leicester, UK; University of Leuven, Belgium; University of Stockholm, Sweden; Utah State University, USA. A portion of this work was carried out at the Jet Propulsion Laboratory, California Institute of Technology, under a contract with the National Aeronautics and Space Administration. 

\end{acknowledgement}

\bibliographystyle{aa}
\bibliography{miri_com}

\end{document}